\documentstyle[11pt,amsfonts]{article}

\newcommand{\be}{\begin{equation}}
\newcommand{\ee}{\end{equation}}
\newcommand{\bea}{\begin{array}}
\newcommand{\ea}{\end{array}}
\newcommand{\beqa}{\begin{eqnarray}}
\newcommand{\eeqa}{\end{eqnarray}}
\newcommand{\bean}{\begin{eqnarray*}}
\newcommand{\eean}{\end{eqnarray*}}

\def\up#1{\leavevmode \raise.16ex\hbox{#1}}

\def\sqr#1#2{{\vcenter{\vbox{\hrule height.#2pt
        \hbox{\vrule width.#2pt height#1pt \kern#1pt
          \vrule width.#2pt}
        \hrule height.#2pt}}}}

\def\BI{{\rm 1\!l}}

\newcommand{\gapproxeq}{\lower .7ex\hbox{$\;\stackrel{\textstyle
>}{\sim}\;$}}
\newcommand{\lapproxeq}{\lower .7ex\hbox{$\;\stackrel{\textstyle
<}{\sim}\;$}}

\newcounter{appendice}

\def\thebibliography#1{{\bf REFERENCES\markboth
 {REFERENCES}{REFERENCES}}\list
 {[\arabic{enumi}]}{\settowidth\labelwidth{[#1]}\leftmargin\labelwidth
 \advance\leftmargin\labelsep
 \usecounter{enumi}}
 \def\newblock{\hskip .11em plus .33em minus -.07em}
 \sloppy
 \sfcode`\.=1000\relax}

\def\BI{{\rm 1\!l}}
\def\up#1{\leavevmode \raise.16ex\hbox{#1}}

\def\sqr#1#2{{\vcenter{\vbox{\hrule height.#2pt
        \hbox{\vrule width.#2pt height#1pt \kern#1pt
          \vrule width.#2pt}
        \hrule height.#2pt}}}}

\def\BI{{\rm 1\!l}}

\begin{document}
\begin{flushright}
\baselineskip=12pt

DIAS-STP-06-17\\
SINP/TNP/06-29\\
\hfill{ }\\
November 2006
\end{flushright}

\centerline{ \LARGE Noncommutative BTZ Black Hole and Discrete Time}

\vskip 2cm

\centerline{ {\sc  B.P. Dolan$^{a}$, Kumar S. Gupta$^{b}$ and
A. Stern$^{c}$ }  }

\vskip 1cm
\begin{center}
 {\it a) Dept. of Mathematical Physics,\\ National University of
   Ireland, Maynooth\\ and Dublin Institute for Advanced Studies,\\ 10,
   Burlington Rd., Dublin, Ireland\\}
 {\it b) Theory Division, Saha Institute of Nuclear Physics\\ 1/AF
   Bidhannagar, Calcutta - 700064, India\\}
 {\it c) Department of Physics, University of Alabama,\\ Tuscaloosa,
Alabama 35487, USA\\}
\end{center}
\vskip 2cm

\vspace*{5mm}

\normalsize
\centerline{\bf ABSTRACT}
We search for all  Poisson brackets for  the BTZ black hole  which are
consistent with the geometry of the
commutative solution and are of lowest order in the embedding coordinates.  For arbitrary values for the angular momentum
we obtain two two-parameter families of contact structures.  We obtain
the symplectic leaves, which characterize the irreducible
representations of the noncommutative theory.  The requirement that
they be invariant under the action of the isometry group restricts to  ${\mathbb{R}}\times S^1$
symplectic leaves, where  ${\mathbb{R}}$ is associated with the Schwarzschild
time.   
Quantization may then lead to a discrete spectrum for the
time operator. 

 \vskip 2cm \vspace*{5mm}

\newpage

Most approaches to noncommutative gravity have involved
deforming the commutative Einstein equations.\cite{Madore:1993br}
-\cite{Balachandran:2006qg}    These approaches range
from simply replacing point-wise products by Moyal star products  in
the Einstein-Hilbert action to the approach  of Aschieri,
et. al.\cite{Aschieri:2005yw} which preserves the
diffeomorphism invariance of general relativity.  Given
the ambiguities of the different schemes, it may be useful to look at other
strategies towards noncommutative gravity.  The one we have in mind here starts with   solutions to the
commutative Einstein equations, with the goal of finding their
noncommutative analogues.   As a first step, one can  try writing down 
Poisson brackets which are consistent with the geometry of some
classical solution.  The noncommutative counterpart of the 
solution is then obtained by `quantization'.  (Here, of course, the
deformation parameter  is the noncommutativity parameter $\theta$ and not $\hbar$.)    The Poisson brackets and
resulting noncommutative algebra may not be unique.  In this regard,
it may be desirable to impose the restriction that the isometry of the classical solution
survives  quantization and is implementable in any irreducible representation of the noncommutative algebra.
An example of this possibility is studied here.

The example is  the three-dimensional (BTZ) black hole
solution, which is characterized by its mass $M$ and angular momentum
$J$, and is asymptotically  AdS${}^3$.\cite{Banados:1992wn},\cite{Carlip:1995qv}   The
BTZ black hole geometry is known to be  the quotient space of the
universal covering space of AdS${}^3$ by some elements of its
 isometry group $SO(2,2)$.  The quotienting breaks the isometry group
 to a two-dimensional subgroup ${\cal G}_{BTZ}$.  Poisson structures (or more precisely,
contact structures) can be obtained for the BTZ black hole which respect
this quotienting and are invariant under ${\cal G}_{BTZ}$. (Some contact structures have already been previously suggested in
\cite{Bieliavsky:2005tj}.)  Here we search for all such Poisson bracket
which are of lowest order  with
 respect to the four-dimensional embedding coordinates for
 AdS${}^3$.  For generic values of the angular momentum, the
 allowable Poisson brackets form two two-parameter  families, and  are  quadratic with
 respect to four-dimensional embedding coordinates.   Depending on the
 values of the parameters, the symplectic leaves (surfaces on which a
 symplectic two-form can be defined) are topologically either
a) ${\mathbb{R}}^2$,  b)   ${\mathbb{R}}\times
{\mathbb{R}}_+$ or c) ${\mathbb{R}}\times S^1$.  The symplectic leaves
 characterize the irreducible representations of the corresponding
noncommutative algebra.  ${\cal G}_{BTZ}$
acts nontrivially in  the case of a)
and b), in general inducing a map between different symplectic
leaves.  It thus transforms between different irreducible
representations of the noncommutative theory, and   ${\cal
  G}_{BTZ}$ cannot be 
implemented as  inner transformations.  On the other hand,  ${\cal
  G}_{BTZ}$ leaves c) invariant and hence also  the corresponding  irreducible 
representations.  So only case c) remains if we impose the restriction
that the isometry of the classical solution survives  quantization.   Moreover,  quantization of the commutative algebra
on ${\mathbb{R}}\times S^1$, where  ${\mathbb{R}}$ corresponds to the
time,  is known to lead to a discrete spectrum for the time operator.
 We speculate that a similar conclusion can be drawn  for c).\footnote{ The 
  possibility of a discrete spectrum for the time in this setting was
  first suggested to us by A.P. Balachandran.}  For us, not all
cylinders need to have a space-time signature, and the coordinate
associated with  ${\mathbb{R}}$  may or may
not be a time-like coordinate.  
Quantum theories on the noncommutative space-time
cylinder have been previously
studied.\cite{Chaichian:2001nw},\cite{Balachandran:2004yh} Other novel
results were shown in
addition to the discrete time spectrum, which   may also  apply here.  Among
them is  the result that time-independent
Hamiltonians are
conserved only up to modulo $2\pi/\theta$.  

In what follows, after  briefly reviewing the geometry of the
commutative BTZ solution and the  quotient space construction,
we write down the two-parameter  families of Poisson brackets and
map them to the  Schwarzschild-like
coordinates and obtain the symplectic leaves.   There is an analogous
quotient space construction in the noncommutative theory which will
be discussed in a later article.\cite{us1}

In terms of Schwarzschild-like
coordinates $(r,t,\phi)$ the invariant measure for the BTZ black
hole is expressed as\cite{Banados:1992wn},\cite{Carlip:1995qv} 
\be ds^2= \biggl( M - \frac {r^2}{\ell^2} - \frac{J^2}{4r^2}\biggr)
dt^2  + \biggl( -M + \frac {r^2}{\ell^2} +
\frac{J^2}{4r^2}\biggr)^{-1} dr^2+ r^2 \biggl(d\phi - \frac J{2r^2} dt\biggr)^2
\;,\label{scwzmtrc}\ee $$0\le r<\infty\;, \;-\infty
< t<\infty\;, \;0\le \phi<2\pi\;,$$
where   $M$ and
 $J$ are the mass and spin, respectively, and   $\Lambda= -1/\ell^2$ is the cosmological constant.  For  $0<|J|<M\ell$, there are two
horizons, the outer and inner horizons,
corresponding respectively to $r=r_+$  and  $r=r_-$, where
\be r_\pm^2 =\frac {M\ell^2}2 \biggl\{ 1\pm \bigg[ 1 - \biggl(\frac
J{M\ell}\biggr)^2\biggr]^{\frac 12} \biggr\}  \;\label{rpm}\ee
 The two horizons
coincide in the extremal case $|J|=M\ell> 0$, while the inner one disappears   for $J=0$, $M>0$.  
The metric is diagonal in the coordinates $(\chi_+,\chi_-,r)$, where
\be \chi_\pm=\frac {r_\pm}{\ell} t- r_\mp \phi \;,\label{defchi}\ee
 \be
ds^2 = \frac{- (r^2-r_+^2) d\chi_+^2 + (r^2-r_-^2)
  d\chi_-^2 }{r_+^2-r_-^2} + \frac{ \ell^2 r^2
dr^2}{(r^2-r_+^2)(r^2-r_-^2)}\;, \label{mtrccpcmr}\ee
which shows that $\chi_+$ is the time-like coordinate in the region   I)  $r\ge
r_+$, $r$ is the time-like coordinate in the region II)   $r_-\le r\le
r_+$ and $\chi_-$ is the time-like coordinate in the region III)  and  $0\le r\le r_-$.

 It was shown that the manifold  of the
BTZ black hole solution is  the quotient space of the
universal covering space of AdS${}^3$ by some elements of the group of
isometries of AdS${}^3$.  The connected component of the latter is  $SO(2,2)$.  
Say AdS${}^3$ is spanned by coordinates $(t_1,t_2,x_1,x_2)$
parameterizing ${\mathbb{R}}^4$, satisfying
\be -t_1^2 -t_2^2 +x_1^2 +x_2^2 =-\ell^2\;\label{tottxoxt}\ee
  Alternatively, one 
can introduce 
 $2\times 2$ real unimodular  matrices\be g=\frac 1{\ell}\pmatrix{t_1+x_1 & t_2
   +x_2\cr -t_2 +x_2 & t_1 -x_1 \cr}\;,\qquad {\rm det} g
 =1\;,\label{gntrmstx}\ee belonging to
   the defining representation of $SL(2,R)$.  The
isometries correspond to the left and right actions on $g$,
\be g\rightarrow h_Lgh_R\;,\qquad    h_L,h_R\in SL(2,R)\label{ismtry}\ee   Since  $( h_L,h_R)$
and  $(- h_L,-h_R)$ give the same action, the connected component of the
isometry group for AdS${}^3$ is  $SL(2,R)\times SL(2,R)/Z_2\approx
SO(2,2)$.

The BTZ black-hole is obtained by discrete identification of points on
the universal covering space   of $AdS_3$.  This   insures 
  periodicity in $\phi$,  $\phi\sim\phi + 2\pi$.  The condition is 
\be g\sim \tilde h_Lg\tilde h_R\;,\label{qtntng}\ee  where $(\tilde
h_L,\tilde h_R)$ are certain elements of $SO(2,2)$. $\tilde h_L$ and
$\tilde h_R$  can be expressed as
 diagonal  $SL(2,R)$ matrices
\be\tilde h_L =\pmatrix{ e^{\pi (r_+-r_-)/\ell} &\cr & e^{-\pi
      (r_+-r_-)/\ell}\cr}\;,\qquad \tilde h_R =\pmatrix{ e^{\pi (r_++r_-)/\ell} &\cr & e^{-\pi
      (r_++r_-)/\ell}\cr}\;\ee 
For $0<|J|<M\ell$, the universal covering space   of $AdS_3$   is covered
by three types of
coordinate patches which are bounded by the two horizons at $r=r_+$ and
$r=r_-$.  For all three coordinate patches, $g$ can be  decomposed  according to
\be g
=\pmatrix{e^{\frac 1{2\ell}(\chi_+-\chi_-)}   & \cr & e^{-\frac 1{2\ell}(\chi_+-\chi_-)}\cr} g^{(0)}(r)\pmatrix{e^{\frac 1{2\ell}(\chi_++\chi_-)}   & \cr & e^{-\frac 1{2\ell}(\chi_++\chi_-)}\cr} \label{gnrlmptsl}\;,\ee
where $g^{(0)}(r)$ is an  $SO(2)$
  matrix which only depends on $r$ and the coordinate patch. 
 The
  periodicity condition for $\phi$  easily follows from (\ref{qtntng}).  
The identification (\ref{qtntng}) breaks the $SO(2,2)$ group of
isometries  to a
two-dimensional subgroup ${\cal G}_{BTZ}$, consisting of only the diagonal matrices in
$\{h_L\}$ and $\{h_R\}$.   ${\cal G}_{BTZ}$ is the isometry group of
the BTZ black hole, and from (\ref{gnrlmptsl}) is associated with
translations in $\chi_+$ and $\chi_-$, or equivalently $t$ and $\phi$,
on $r=$constant surfaces.

For generic   spin, $0<|J|< M\ell$ (and   $M>0$), we shall  search for   Poisson
brackets for the  matrix elements of $g$ which are  polynomial of  lowest
order.  They  should be
consistent with the quotienting (\ref{qtntng}), as well as the
unimodilarity condition and, of course, the Jacobi identity.  
  For convenience we write the $SL(2,R)$ matrix as
\be g=\pmatrix{\alpha & \beta\cr\gamma & \delta\cr}\;,\qquad
\alpha\delta -\beta\gamma=1\;,\label{gabgd}\ee Under  the  quotienting
  (\ref{qtntng}):
\beqa \alpha & \sim & e^{\;\;2\pi r_+ /\ell}\; \alpha \cr
 \beta & \sim & e^{-2\pi r_- /\ell}\; \beta \cr
 \gamma & \sim & e^{\;\;2\pi r_- /\ell}\; \gamma \cr
 \delta & \sim & e^{-2\pi r_+ /\ell}\; \delta \label{lmntsqtntng}
 \eeqa All quadratic
  combinations of matrix elements
  scale differently, except for
 $\alpha\delta$ and $\beta\gamma$, which are invariant under (\ref{lmntsqtntng}).  Lowest order
 polynomial expressions for  the Poisson brackets of
 $\alpha,\beta,\gamma$ and $\delta$ which are preserved
 under (\ref{lmntsqtntng}) are quadratic and  have the
 form
\be \left. \matrix{\{\alpha,\beta\}=c_1 \alpha\beta\; &\;
  \{\alpha,\gamma\}=c_2 \alpha\gamma \;&\;  \{\alpha,\delta\}=f_1(
  \alpha\delta,\beta\gamma) \cr\{\beta,\delta\}=c_3 \beta\delta \;&\;
  \{\gamma,\delta\}=c_4 \gamma\delta \; &\; \{\beta,\gamma\}=f_2(
  \alpha\delta,\beta\gamma) \cr} \right.\;, \label{pbitoxm}\ee  
where $c_{1-4}$ are constants and $f_{1,2}$ are
functions.\footnote{More generally, if
  we drop the assumption that the Poisson brackets  are  polynomial of  lowest
order we can replace the constants $c_i$ by functions of
$\alpha\delta$ and $ \beta\gamma$.  So  for example,
$\{\alpha,\beta \}=p_1(\alpha\delta, \beta\gamma)\alpha\beta$,
where $p_1$ is an arbitrary function.   These brackets, in
general, will have more complicated
transformation properties under the action of $SO(2,2)$.}  They are
constrained by 
\beqa c_1+c_2&=&c_3+c_4 \cr f_1(
  \alpha\delta,\beta\gamma)&=& (c_1+c_2)\beta\gamma\cr f_2(
  \alpha\delta,\beta\gamma)&=&(c_2-c_4)\alpha\delta\;,\label{foneftwo} \eeqa after demanding
that det$g$ is a Casimir of the algebra.
From (\ref{pbitoxm}) there are three independent constants  $c_{1-4}$.
 Further restrictions on the constants   come from the Jacobi identity, which leads to the
following two possibilities:

$$ {\rm A.} \quad c_2=c_4 \qquad{\rm and}\qquad {\rm B.} \quad  c_2=-c_1$$

\noindent Both cases  define two-parameter families
  of Poisson brackets.  Say we call $c_2$ and $c_3$ the two independent parameters.
 The two cases are connected by an $SO(2,2)$ transformation.  Case A goes to case B  when
\be g=\pmatrix{\alpha & \beta\cr\gamma & \delta\cr}\rightarrow
  g'= \pmatrix{\beta & -\alpha\cr\delta &-\gamma \cr}=g
 h^{(0)}_R\;,\qquad  h^{(0)}_R= \pmatrix{ &-1\cr 1 &\cr}\;,
\ee along with \be c_3\rightarrow
  c_2
\qquad c_2\rightarrow c_3
\ee  In terms of the embedding coordinates, this corresponds to
 $(t_1,t_2,x_1,x_2)\rightarrow (t_2,-t_1,x_2,-x_1)$.

  There are three types of coordinate patches in
the generic case of $M>0$ and $0<|J|< M\ell$, and their boundaries
are the two horizons.   Denote them again by:   I)  $r\ge
r_+$, II)  $r_-\le r\le r_+$ and  III)  $0\le r\le r_-$.  The corresponding maps to
$SL(2,R)$ are given by (\ref{gnrlmptsl}), with 

\noindent I)  $r\ge r_+$, 
\be g^{(0)}(r)= g^{(0)}_I(r) =\frac 1{\sqrt{r_+^2 - r_-^2}} \pmatrix{\sqrt{{r^2 - r_-^2}} & \sqrt{r^2-r^2_+}\cr  \sqrt{r^2-r^2_+} & \sqrt{{r^2 - r_-^2}} \cr}\;\label{mpscldsl1} \ee

\noindent II)  $r_-\le r\le r_+$,  
\be g^{(0)}(r)= g^{(0)}_{II}(r) =\frac 1{\sqrt{r_+^2 - r_-^2}} \pmatrix{\sqrt{{r^2 - r_-^2}} & -\sqrt{r^2_+-r^2}\cr  \sqrt{r^2_+-r^2} & \sqrt{{r^2 - r_-^2}} \cr}\;\label{mpscldsl2} \ee

\noindent III)  $0\le r\le r_-$,  
\be g^{(0)}(r)= g^{(0)}_{III}(r) =\frac 1{\sqrt{r_+^2 - r_-^2}} \pmatrix{\sqrt{{r_-^2 - r^2}} & -\sqrt{r^2_+-r^2}\cr  \sqrt{r^2_+-r^2} & -\sqrt{{r^2_- - r^2}} \cr}\;\label{mpscldsl3} \ee
  Using the maps (\ref{mpscldsl1}-\ref{mpscldsl3}), we can write the Poisson
  brackets for the various cases in terms of the   Schwarzschild-like
coordinates $(r,t,\phi)$.  The results are the same in all three
  coordinate patches.   For the two-parameter families  A and B one gets:

\indent A.
\beqa \{\phi,t\}&=& \frac {\ell^3}2\; \frac{c_3-c_2}{r_+^2 - r_-^2}
\cr  \{r,  \phi\} &=&-\frac{ {\ell r_+}(c_3+c_2)} {2r}\;\frac{r^2 - r_+^2}{r_+^2 - r_-^2}
\cr  \{r, t \} &=& -\frac  {\ell^2 r_-(c_3+c_2)} {2r}\;\frac{r^2 - r_+^2}{r_+^2 - r_-^2}\;
\label{pbscA}\;, \eeqa

\indent B.
\beqa \{\phi,t\}&=& \frac {\ell^3}2\; \frac{c_3-c_2}{r_+^2 - r_-^2}
\cr  \{r, \phi\} &=&  - \frac{{\ell r_-}(c_2+c_3)} {2r}\;\frac{r^2 - r_-^2}{r_+^2 - r_-^2}
\cr  \{r, t \} &=&- \frac{{\ell^2 r_+}(c_2+c_3)} {2r}\;\frac{r^2 - r_-^2}{r_+^2 - r_-^2} \label{pbscB}
\eeqa  These Poisson brackets are invariant under the action of  the
isometry group   ${\cal G}_{BTZ}$ of
the BTZ black hole. The first bracket agrees in both cases.   The latter two  brackets 
vanish at the outer
horizon $r=r_+$ for case A, and the inner horizon $r=r_-$ for case B.
A central element of the Poisson  algebra can be
constructed out of the Schwarzschild coordinates for both cases.  It
is given by
\be \rho_\pm =(r^2-r_\pm^2) \exp{\biggl\{- \frac{2\kappa\chi_\pm}\ell}\biggl\}\;,\qquad c_2\ne c_3\;,\label{rhominus}\ee 
where  the upper
and lower sign correspond to case A and B, respectively,
\be \kappa =  \frac
  {c_3+c_2}{c_3-c_2}\;,\ee and $\chi_\pm$ were defined in
  (\ref{defchi}).  The  $\rho_\pm=$constant surfaces define
symplectic leaves, which are topologically  ${\mathbb{R}}^2$ for
generic values of the parameters (more specifically,  $ c_2\ne \pm
c_3$).  We can coordinatize them by $\chi_+$ and $\chi_-$.  One then
has a trivial Poisson algebra in the coordinates $(\chi_+,\chi_-,\rho_\pm)$:
\be \{\chi_+,\chi_-\} =\frac{\ell^2}2(c_3-c_2)   \qquad
\{\rho_\pm,\chi_+\}=\{\rho_\pm,\chi_-\}=0\ee  The  action of  the
  ${\cal G}_{BTZ}$ transforms one 
 symplectic leaf to another, except for the case $c_2= -c_3$ which we
 discuss later. 

The above can be readily extended to the case of zero angular momentum
by simply setting $r_-=0$.
The region  III is then absent in this case.\footnote{Zero angular momentum
also allows for Poisson brackets which are linear with respect to the
four dimensional embedding coordinates  and consistent with
(\ref{qtntng}).  This  will
be discussed in a later article.\cite{us1}} 
  On the other hand, the Poisson brackets  (\ref{pbscA}) and (\ref{pbscB}) are undefined in the extremum
case $J=M\ell$, or $r_+=r_-$, for finite coefficients $c_i$.\footnote{In the extremal case, $h_L$ reduces to
the identity, and at first glance it appears that more general Poisson
brackets than   (\ref{pbitoxm}) and (\ref{foneftwo}), and consequently
 (\ref{pbscA}) and (\ref{pbscB}), are admissible.   This is because the products  $\alpha\beta$, $
\alpha\delta$, $  \gamma\delta$ and $\beta\gamma$
 are unaffected by the  quotienting.  Thus the quotienting conditions allows one to
generalize  (\ref{pbitoxm}) such that $\{\alpha,\beta\}$, $
\{\alpha,\delta\}$, $  \{\gamma,\delta\}$ and $\{\beta,\gamma\}$
  depend on those four products.  But the system reduces to (\ref{pbitoxm}) and (\ref{foneftwo}) after demanding that 
 det$g$ is a Casimir of the algebra.}  The
brackets, however,  may be rendered finite by
first considering $J<M\ell$ with the coefficients  $c_i$ proportional to ${r_+^2 - r_-^2}
$ and then taking the limit $J\rightarrow M\ell$.

In passing to the noncommutative theory, the operator 
associated with $\rho_\pm$ is central in the quantum algebra and
  proportional to the identity in any  irreducible
representation.  Irreducible representations then select  $\rho_\pm=$constant
surfaces and  the isometry group  ${\cal G}_{BTZ}$
thus  maps between different
irreducible representations, and thus cannot be implemented as inner transformations.   In any irreducible representation the algebra is generated by  the noncommutative analogues of  $\chi_+$ and $\chi_-$.
The resulting noncommutative theory differs from the Gr\"onewald-Moyal plane
since  $\chi_+$ and
$\chi_-$ are not cartesian coordinates.   After re-writing the commutative metric (\ref{mtrccpcmr}) in terms of coordinates 
 $(\chi_+,\chi_-,\rho_\pm)$
 and restricting to   the  $\rho_\pm=$constant surface
one gets by \be
ds^2|_{ \rho_\pm} = \frac{- (r^2-r_+^2) d\chi_+^2 + (r^2-r_-^2)
  d\chi_-^2 }{r_+^2-r_-^2} + \frac{r^2-r_\pm^2}{r^2-r_\mp^2}
\kappa^2 d\chi_\pm^2
 \ee  As a result, these metric components  will not in general be fixed by the
irreducible representation.  In attempting to write down a
noncommutative field theory in this case, one cannot expect to treat
the metric as a background.  Neither is the signature of the metric  fixed by the
irreducible representation, as
it can differ in different regions
on the surface, which is evident from the  determinant of the
commutative metric $ {\tt g}$ for fixed $\rho_\pm$
\be \det {\tt g}|_{\rho_\pm}=  -(r^2 - r_\pm^2) \biggl(\frac{r^2- r^2_\mp}{r_+^2-r_-^2}\; \mp\;\kappa^2\biggr)
\label{sgntr}\ee
The surface has a  Minkowski signature for $r$ sufficiently large, and space-time
noncommutativity results in the noncommutative theory.  On the other
hand, there may be regions where (\ref{sgntr}) is positive which is then
associated with space-space noncommutativity.

We next discuss the two exceptional cases: $c_2= c_3$ and $c_2=- c_3$.

The above results cannot be applied when
$c_2=c_3$ since $\kappa$, and hence  $\rho_\pm$, are
ill-defined.    Instead, $\chi_\pm$ 
is central in the Poisson algebra in this case, where the upper and
lower sign again correspond to case A and B, respectively.  The
$\chi_\pm=$constant surfaces define the
symplectic leaves, which are topologically  ${\mathbb{R}}\times
{\mathbb{R}}_+$, parametrized by $r$ and \be  \xi_\mp =\pm \frac {r\chi_\mp}{r^2-r_\pm^2}
 \ee 
In terms of these variables, the  Poisson brackets are
\be \{r,\xi_\mp\} = \ell c_2 \qquad \{\chi_\mp,r\}=
\{\chi_\pm,\xi_\mp\}=0\; \ee
 Irreducible representations now select the $\chi_\pm=$constant
 symplectic leaves.  As before, the isometry group  ${\cal G}_{BTZ}$
is a map between different symplectic leaves and hence different
irreducible representations in the quantum theory. 
  The  $\chi_\pm=$constant
 surfaces are characterized by the metric
 \be
ds^2|_{\chi_\pm} = \pm \frac{r^2-r_\mp^2}{r_+^2-r_-^2}\biggl(\frac1r (r^2-r_\pm^2) d\xi_\mp + \frac 1{r^2} (r^2+r^2_\pm )\xi_\mp
    dr\biggr)^2  + \frac{ \ell^2 r^2
dr^2}{(r^2-r_+^2)(r^2-r_-^2)}
\;, \ee whose determinant is simply
\be \det {\tt g}|_{\chi_\pm}=\pm\ell^2\; \frac{r^2 - r_\pm^2}{r_+^2 -r_-^2} 
\ee
For case A, the surface has a  Euclidean signature for $r>r_+$ and
Minkowski signature for $r<r_+$.  For case B, the surface has a  Minkowski signature for $r>r_-$ and
Euclidean signature for $r<r_+$.

The case of $c_2 =-c_3$ is the intersection of   case A and B.
Here $\kappa$ vanishes  and from (\ref{rhominus}), the radial coordinate is in the center of,
the algebra.     $r=$constant define ${\mathbb{R}}\times S^1$
 symplectic leaves, and they are invariant under the  action of  ${\cal G}_{BTZ}$.
 The coordinates  $\phi$ and $t$ parametrizing any such surface are  canonically
conjugate:  \be \{\phi,t\}= \frac {c_3 \ell^3}{r_+^2-r^2_-}\qquad \{\phi_\pm,r\}=\{t,r\}=0 \label{ctemcth}\ee The Poisson brackets 
can be interpreted in terms of a twist\cite{Connes:2000tj} in the decomposition of $g$
given in (\ref{gnrlmptsl}), where the twist is with
respect to the first and third matrices.      
In passing to the noncommutative theory, we need to define a
deformation of the commutative algebra generated by  $t,e^{i \phi}$ and $r$.   Call
 the corresponding quantum operators $\hat t, e^{i\hat\phi}$ and $\hat r$, respectively.
Their commutation relations are\footnote{It should be noted  that
  there exits a  quantization ambiguity associated with the set of allowed  canonical
  transformations on the cylinder.  So for example, the commutation
  relations are unchanged under the
  redefinitions  $$\hat t
\rightarrow \hat t' =\hat t + F_1(\hat r)\qquad e^{i\hat
  \phi}\rightarrow e^{i\hat \phi'}=e^{i\{\hat \phi+ F_2(\hat r)\hat
  t\}}\;,$$  for arbitrary functions $F_1$ and $F_2$.  More
 physical input is needed to resolve this quantization ambiguity.}
\be [e^{i\hat \phi},\hat t] = \theta e^{i\hat \phi}\qquad [\hat r,
\hat t] =  [\hat r,e^{i\hat \phi}] =0\;,\label{qntmalg} \ee
where from (\ref{ctemcth}) the constant $\theta$ is linearly
related to $\ell^3/(r_+^2-r_-^2)$.  There are now two central elements
in the algebra: i) $\hat r$ and ii)   $e^{-2\pi i\hat
  t/\theta}$.  From i),  irreducible
representations  select the ${\mathbb{R}}\times S^1$
symplectic leaves.  Unlike in all the previous cases,  the action of
${\cal G}_{BTZ}$ does not take you out of any particular irreducible
representation, and in this sense we can say that the isometry of the
classical solution survives  quantization.   The action of ${\cal
  G}_{BTZ}$ can be implemented 
with inner transformations.  Say $X_t$ and $X_\phi$ are Killing vectors
generating translations in $t$ and $\phi$, respectively.  They act on
 functions $\hat A$ on the noncommutative space according to
\be   X_t \hat A =-\frac1\theta [\hat\phi, \hat A] \qquad  X_\phi
\hat A =\frac1\theta [\hat t, \hat A] \ee
  The determinant of the metric for any symplectic leaf is a function of $r$, 
\be \det {\tt g}|_{r}=-\frac 1{\ell^2} \;(r^2 - r_+^2)(r^2 -r_-^2) 
\;,\ee and thus the signature, as
well as the commutative metric, are fixed by the irreducible representation. The
cylinders have a  Minkowski signature for regions I and III, and a
Euclidean  signature for region II. 
With regard to the central element ii)
  $e^{-2\pi i\hat
  t/\theta}$, one can  identify it with   $e^{i\chi}\BI$ in an irreducible representation. 
The spectrum of $\hat t$ is then
discrete\cite{Chaichian:2001nw},\cite{Balachandran:2004yh}  \be n\theta
-\frac {\chi\theta}{2\pi}\;,\;\;n\in
{\mathbb{Z}}\label{dsctsptmt}\ee
 In associating $\hat t$ with the Schwarzschild coordinate $t$, we
 recall that the latter is the time for the
exterior of the black hole, but not for the interior.  More precisely,  $X_t$ is
time-like provided $r>r_{erg}$,  where   $r_{erg}$ is the radius  of the ergosphere
(or ergocircle),
 $r_{erg}^2 = r_+^2 + r_-^2$.

Although the Poisson
brackets (\ref{pbscA}) and  (\ref{pbscB}) are  invariant
under  the action of the  isometry group of the black hole, they  are not
invariant under the larger group of   $SO(2,2)$ transformations (\ref{ismtry}).  On the other hand,  Poisson structures  can be  consistently assigned to  $SO(2,2)$  
  such that it  defines a Lie-Poisson group and
(\ref{ismtry}) defines a   Poisson map.  The $SO(2,2)$ group  get
q-deformed upon quantization.  The noncommmutative BTZ black hole can 
be obtained from this quantum group by  quotenting   in a manner analogous
to  (\ref{qtntng}).
This, along with an  attempt at field theory on the noncommutative background,  will
be pursued in  later articles.\cite{us1}

\subsubsection*{Acknowledgements}
We are very grateful to A.P. Balachandran, A. Pinzul and  P.~Presnajder for  
useful discussions.  We also thank  P.~Presnajder for his hospitality
during a stay at Dept. of Physics, Comenius University, Bratislava,
where this work originated.

\end{document}